\documentclass{iopjournal}

\usepackage[utf8]{inputenc}
\usepackage{textgreek}
\usepackage{ragged2e}
\usepackage{siunitx}
\usepackage{booktabs}
\usepackage{multirow}
\usepackage{amsmath,amssymb}

\begin{document}
\articletype{Paper}

\title{Investigation of the Microstructural and Magnetic Properties of Ni on GeSn Layers}

\author{Tharundev VV$^1$\orcid{0009-0005-0313-8907}, Dushyant Singh$^1$\orcid{0000-0002-0386-1573}, Subha Maity$^1$\orcid{0009-0007-6948-9552}, Vigneshwari S R$^1$\orcid{0009-0000-4036-4456}, Dhammapriy Gayakwad$^2$\orcid{0009-0007-6357-7228}, Shivesh Yadav$^3$\orcid{0000-0003-0897-9463} and Krista R Khiangte$^{1,*}$\orcid{0000-0003-0294-7815}}

\affil{$^1$Department of Physics, Indian Institute of Technology Gandhinagar, Gujarat 382424, India}
\affil{$^2$Department of Physics, Indian Institute of Technology Bombay, Mumbai 400076, India}
\affil{$^3$Department of Condensed Matter Physics and Material Science, Tata Institute of Fundamental Research, Mumbai 400005, India}
\affil{$^*$Author to whom any correspondence should be addressed.}
\email{krista.khiangte@iitgn.ac.in}


\begin{abstract}
Thermal processing of nickel (Ni) layers on germanium-tin (GeSn) can strongly modify the Ni film and the Ni/GeSn interface within a restricted thermal budget; however, the influence of GeSn crystalline quality on Ni-Ge-Sn interfacial evolution and the magnetic response of the Ni layer remains incompletely understood. Here, Ni (15~nm)/Ge$_{1-x}$Sn$_x$/Ge/Gd$_2$O$_3$/Si(111) heterostructures ($x = 0.08$ and $0.10$) were investigated following rapid thermal annealing at 573 and 973~K for 10~s using X-ray diffraction, pole-figure analysis, X-ray photoelectron spectroscopy depth profiling, and temperature- and field-dependent magnetization measurements. Annealing at 573~K preserves the crystallographic texture of Ge$_{0.92}$Sn$_{0.08}$, whereas Ge$_{0.90}$Sn$_{0.10}$ exhibits microtwin formation and reduced structural integrity. Annealing at 973~K, employed as a severe upper-bound thermal reference, causes extensive structural degradation in both heterostructures, with substantially greater deterioration in the higher-Sn sample. XPS depth profiling reveals progressive Ni-Ge-Sn intermixing accompanied by pronounced Sn surface segregation with increasing annealing temperature. Magnetization measurements show coexisting ferromagnetic and superparamagnetic contributions in the as-deposited and 573~K samples, whereas annealing at 973~K strongly suppresses the ferromagnetic response, producing highly unsaturated $M(H)$ loops together with reduced saturation magnetization and coercivity. These magnetic changes are consistent with extensive interfacial alloying and dilution of the metallic Ni layer. This study establishes a direct correlation between GeSn crystalline quality, interfacial diffusion, and magnetic behavior.
\end{abstract}

\section{Introduction}
Traditional complementary metal-oxide-semiconductor (CMOS) technology is approaching its fundamental scaling limits due to quantum confinement effects, increased parasitic losses, and process-related constraints, which hinder further improvements in high-speed data processing and transmission \cite{bravaix_cmos_2023,islam_challenges_2026}. To address these challenges, the integration of optoelectronic functionalities with the silicon platforms has emerged as a promising solution \cite{shekhar_roadmapping_2024,chou_highdetectivity_2025}. However, the realization of fully integrated silicon photonics is critically limited by the lack of an efficient, silicon-compatible light source. Germanium–tin (GeSn), a group-IV alloy, has recently attracted significant attention as a potential candidate for silicon-based optoelectronic devices\cite{cui_high-power_2025}. By increasing the Sn content, the band structure of GeSn can be engineered to undergo an indirect-to-direct bandgap transition at Sn concentrations exceeding 8-9 at.\%, enabling efficient light emission \cite{wirths_lasing_2015,dev_wafer-scale_2020}. Despite these advantages, the growth of high-quality GeSn alloys remains challenging due to the inherently low equilibrium solubility of Sn in Ge ($<1\%$), the large lattice mismatch between Ge and α-Sn (~14\%), and the low eutectic temperature of the Ge–Sn system (~239.1 °C) \cite{gayakwad_epitaxial_2023}. These factors promote Sn segregation and phase separation at elevated temperatures, leading to structural instability. To overcome these limitations, non-equilibrium growth techniques, such as Molecular Beam Epitaxy (MBE), Chemical Vapor Deposition (CVD), and radio-frequency (RF) sputtering have been widely employed to stabilize high Sn incorporation. Several studies have reported the successful growth of GeSn alloys with Sn concentrations exceeding 20 at.\% while maintaining reasonable crystalline quality \cite{zhang_gesn_2026,shibayama_comprehensive_2024,liu_defect-free_2016}.

For the realization of high-performance GeSn-based electronic and optoelectronic devices, the formation of a stable, low-resistance metal contact is essential. However, achieving reliable contacts in GeSn is inherently challenging due to the metastable nature of Sn within the Ge lattice. During metallization and subsequent thermal processing, Sn tends to segregate toward the metal/GeSn interface, leading to non-uniform interfacial reactions, morphological roughening, and phase inhomogeneity \cite{segura-ruiz_dynamics_2025,m_eldose_strain-mediated_2023,cai_thickness-dependent_2022}. These effects collectively degrade the interface quality and result in an increased specific contact resistance. Among the various contact materials, nickel (Ni) has emerged as a promising candidate due to its established compatibility with CMOS technology and its ability to form low-resistivity phases in group-IV semiconductors \cite{li_electrical_2014,jeon_formation_2018}. In Ge-based systems, Ni readily reacts to form NiGe, which exhibits a low sheet resistance, good thermal stability, and favorable scaling behavior \cite{jablonka_formation_2017}.

In conventional CMOS processing, such reactions are typically achieved using Rapid Thermal Annealing (RTA), which enables controlled phase formation through short-duration, high-temperature cycles while limiting excessive diffusion. This approach is particularly important for maintaining sharp interfaces and minimizing unwanted elemental redistribution \cite{ko_formation_2006,kittl_silicides_2008}. In the context of GeSn, the role of RTA becomes even more critical due to the low thermal stability of the alloy. The short thermal budgets associated with RTA provide a pathway to initiate Ni-induced reactions while mitigating Sn segregation and preserving the alloy composition \cite{xu_time-_2025}. Extending this approach to GeSn, recent studies suggested that Ni can form a ternary Ni–Ge–Sn intermetallic phase at relatively low temperatures (~573–673~K) \cite{khelidj_nickel_2024}, which is consistent with the thermal constraints imposed by the low eutectic temperature of the Ge–Sn system (~512.1~K) \cite{gayakwad_epitaxial_2023}. Notably, Ni-induced reactions in GeSn can proceed under such constrained conditions, enabling contact formation without severely degrading the material \cite{quintero_ni-based_2023}. The resulting Ni(GeSn) phase has been reported to exhibit sheet resistance values comparable to those of conventional Ni germanides, indicating its potential for achieving low-resistance source/drain contacts in GeSn-based devices \cite{quintero_ni-based_2023}.

Despite extensive research devoted to the Ni-based metallization of GeSn, several aspects of the interfacial reaction process remain incompletely understood. Previous studies have primarily focused on phase formation, contact resistivity and thermal stability, establishing Ni(GeSn) as a low-resistance contact phase for GeSn-based devices \cite{khelidj_nickel_2024,quintero2023use}. However, the microscopic mechanisms governing the Ni–GeSn reaction have received comparatively less attention. In particular, the influence of the structural quality of the GeSn epilayer on interfacial reactions and phase evolution remains largely unexplored. Owing to the metastable nature of GeSn, variations in the crystalline quality, defect density, strain relaxation, and Sn distribution are expected to influence the atomic diffusion pathways and reaction kinetics during thermal processing. Understanding these material characteristics is therefore important for interpreting the thermal evolution of Ni layers on GeSn.

In addition to structural considerations, the evolution of the intrinsic properties of the Ni layer during thermal processing has received little attention. Nickel is a ferromagnetic metal whose magnetic response is highly sensitive to changes in the composition, crystal structure, and local atomic environment. Consequently, interdiffusion, alloy formation, and Sn redistribution at the interface are expected to modify the magnetic properties of the Ni layer. Magnetization measurements can provide information that complements conventional structural and chemical characterization techniques, offering a sensitive probe for interfacial alloying, phase evolution, and degradation of the metallic Ni film.

To date, the magnetic properties of Ni layers on GeSn have remained largely unexplored, and the influence of GeSn crystalline quality on the thermal evolution of the Ni/GeSn stack has not been systematically investigated. In this work, we study Ni/Ge$_{1-x}$Sn$_x$/Ge/Gd$_2$O$_3$/Si(111) heterostructures containing 8 and 10 at.\% Sn using X-ray diffraction, pole-figure analysis, X-ray photoelectron spectroscopy depth profiling, and temperature- and field-dependent magnetization measurements. This study establishes a direct correlation between GeSn crystalline quality, interfacial diffusion, and magnetic behavior, and reveals the critical role of GeSn crystalline quality in determining the thermal stability of the Ni/GeSn heterostructure. Electrical characterization of contact performance is outside the scope of this study.

\section{Experimental Details}
A 35~nm epitaxial Gd$_{2}$O$_{3}$ buffer layer was deposited on a Si(111) substrate using a DCA S1000 multi-chamber MBE system at Leibniz University, Hannover, Germany. Subsequently, GeSn epilayers with an intermediate Ge buffer layer were grown on the Gd$_{2}$O$_{3}$/Si(111) substrates at the IIT Bombay Nanofabrication Facility (IITBNF). Two samples with Sn concentrations of 8~at.\% and 10~at.\% were prepared following the procedure described in Ref.~\cite{tharundevexplicit2026}. Prior to the Ni deposition, the substrates were cleaned sequentially with acetone and isopropanol, followed by a 3~s dip in 2\% HF solution to remove the native oxide layer. A 15~nm Ni layer was then deposited by thermal evaporation at a base pressure of $5 \times 10^{-6}$~mbar. The Ni film thickness was monitored in real time using a quartz crystal microbalance. For direct microstructural characterization of the as-deposited Ni layer, a Ni film was also deposited under the same growth conditions onto a TEM grid.

Post-deposition samples were subjected to RTA in a nitrogen environment at 573~K and 973~K, respectively, for 10~seconds to study the influence of the annealing temperature on the interfacial evolution of the Ni/GeSn stack. Here, RTA at 573~K represents a technologically relevant thermal budget for initiating Ni-GeSn interfacial reactions, whereas RTA at 973~K is employed as a deliberately severe, upper-bound reference condition to map the high-temperature limit of structural and interfacial degradation, rather than as a recommended process window for device fabrication. Structural investigations were conducted utilizing high-resolution X-ray diffraction (XRD) on a Rigaku SmartLab diffractometer, which featured a 9~kW rotating Cu anode source, parabolic mirror optics, and a double-crystal Ge(220) monochromator. For pole figure analysis, the tilt angle ($\chi$) was varied from 0$^\circ$ to 75$^\circ$, and the azimuthal angle ($\phi$) was rotated from 0$^\circ$ to 360$^\circ$. X-ray photoelectron spectroscopy (XPS) depth profiling was conducted using a K-Alpha spectrometer (Thermo Fisher Scientific) with sequential Ar$^+$ ion sputtering to examine the chemical composition and interfacial interdiffusion across the Ni/GeSn heterostructures. The magnetic properties of the samples were investigated through temperature-dependent magnetization, $M(T)$, and magnetic field-dependent magnetization, $M(H)$ measurements using a vibrating sample magnetometer-superconducting quantum interference device (VSM-SQUID) attachment integrated with a Physical Property Measurement System (PPMS).

\section{Results and Discussion}

\subsection{Structural Analysis}
\begin{figure}[t!]
     \centering
     \includegraphics[width=0.9\textwidth]{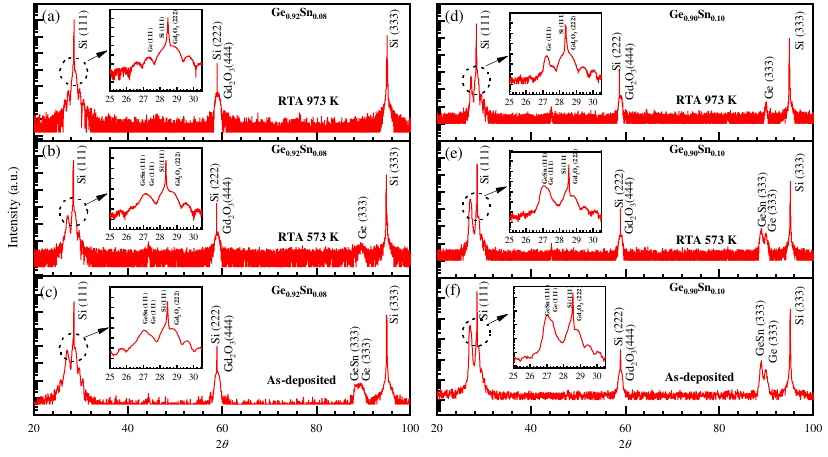}
     \caption{$\theta$-2$\theta$ scans of Ni/\allowbreak Ge(1$-$x)Sn$_x$/\allowbreak Ge/\allowbreak/Gd$_2$O$_3$ Si(111) samples: (a-c) $x = 8\%$ for RTA at 973~K, RTA at 573~K, and as-deposited conditions; (d-f) $x = 10\%$ for RTA at 973~K, RTA at 573~K, and as-deposited conditions.}
     \label{xrd1}
\end{figure}

Fig.~\ref{xrd1} presents the wide-angle $\theta$-2$\theta$ XRD scans of Ni/\allowbreak Ge$_{1-x}$Sn$_x$/\allowbreak Ge/\allowbreak Gd$_2$O$_3$/\allowbreak Si(111) layered structure with Sn concentrations of 8\% (a-c) and 10\% (d-f), corresponding to the as-deposited, 573 K RTA, and 973 K RTA conditions. For both compositions, the GeSn diffraction peak exhibited a systematic shift toward higher diffraction angles upon annealing. This shift indicates a reduction in the lattice parameter, which can be attributed to Sn out-diffusion and segregation during the thermal treatment, leading to partial relaxation of the strained GeSn layer\cite{gayakwad_epitaxial_2023,singh2025growth}. The effect was more pronounced at higher annealing temperatures (973 K), suggesting enhanced thermally activated diffusion processes. Additionally, the evolution of the peak shape and intensity with annealing reflects the modifications in the crystalline quality and interfacial structure. The presence of Pendell\"osung fringes in the as-deposited and low-temperature annealed samples signifies a high-quality, coherent interface with a well-defined thickness and smoothness\cite{bhargava2013,richard2014}. However, these fringes become less distinct or diminish after high-temperature annealing, indicating degradation of interface abruptness, likely due to interdiffusion and possible formation of Ni-Ge-Sn intermetallic phases.

\subsection{Pole-figure measurement}
\begin{figure}[t!]
		\centering
		\includegraphics[width=0.8\textwidth]{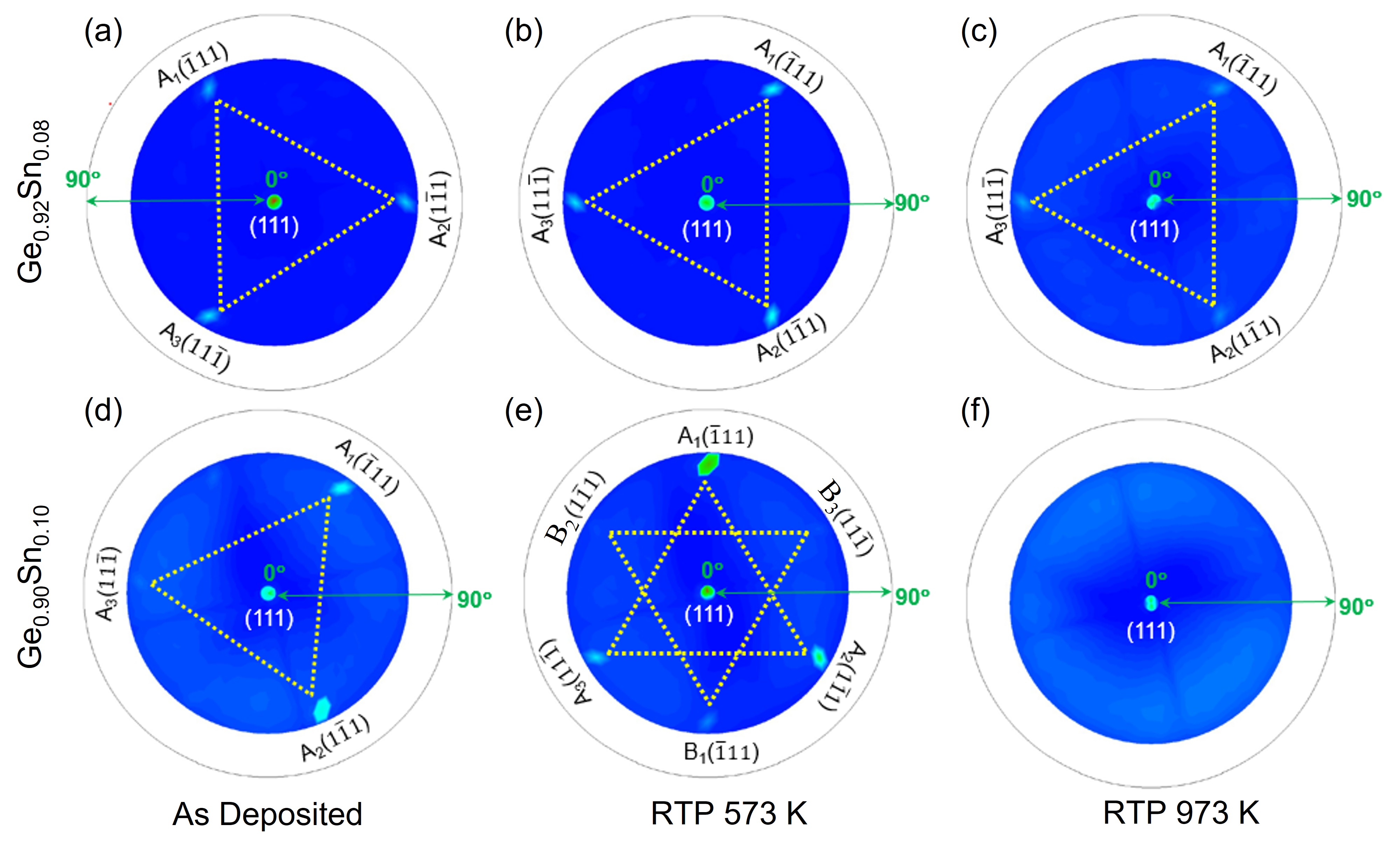}
		\caption{XRD pole figure measurements were carried out at the Bragg angle corresponding to the GeSn(111) reflection. Figures~(a)-(c) present the pole figure measurements for Ge$_{0.92}$Sn$_{0.08}$ in the as-deposited state and after RTA at 573 K and 973 K, respectively. Figures~(d)-(f) show the corresponding measurements for Ge$_{0.90}$Sn$_{0.10}$ under the same conditions.}
		\label{PF}
\end{figure}

To evaluate the structural quality and defect evolution on a global scale, pole figure measurements were performed on Ni/\allowbreak Ge$_{1-x}$Sn$_x$/\allowbreak Ge/\allowbreak Gd$_2$O$_3$/\allowbreak Si(111) with the Bragg angle fixed at the GeSn(111) reflection. Figure~\ref{PF}(a-c) and \ref{PF}(d-f) show the pole figures for the Ge$_{0.92}$Sn$_{0.08}$ and Ge$_{0.90}$Sn$_{0.10}$ films, respectively, under the as-deposited, 573 K RTA, and 973 K RTA conditions. The center of each pole figure corresponds to the zero-tilt condition ($\chi = 0^\circ$), representing the GeSn(111) reflection. For both compositions in the as-deposited state, a triad of poles was observed at $\chi \approx 70.5^\circ$, labeled A$_1$, A$_2$, A$_3$, corresponding to the three $\{1\bar{1}1\}$ planes \cite{singh2025growth}. These poles were separated by $120^\circ$ in the azimuthal direction ($\Delta\phi = 120^\circ$), confirming the expected threefold symmetry of the diamond cubic GeSn lattice along the (111) orientation, and indicating good epitaxial alignment.

Upon annealing at 573 K, the Ge$_{0.92}$Sn$_{0.08}$ sample retained the same triad without the appearance of additional poles, suggesting that the crystalline quality and orientation remained largely preserved at this temperature. In contrast, the Ge$_{0.90}$Sn$_{0.10}$ sample exhibited the emergence of additional poles (labeled B$_1$), which can be attributed to the formation of micro-twin defects \cite{rawat2021epitaxial}. This indicates that a higher Sn content reduces the structural stability and promotes defect formation, even at moderate annealing temperatures. A clear difference between the two compositions is observed after annealing at 973 K. For the Ge$_{0.92}$Sn$_{0.08}$ sample, the triad of poles remained visible, although with reduced intensity, indicating partial degradation but overall retention of the crystallographic texture. In contrast, the Ge$_{0.90}$Sn$_{0.10}$ sample showed a complete disappearance of poles, signifying severe deterioration of the crystalline order, likely due to Sn segregation, phase separation, or amorphization at elevated temperatures.

These results demonstrate that increasing Sn concentration significantly affects the thermal stability of GeSn films. While the lower Sn composition (8\%) maintains its structural integrity even after high-temperature annealing, the higher Sn composition (10\%) is considerably more susceptible to defect formation and structural degradation, highlighting the critical role of Sn content in determining the robustness of the GeSn layers under thermal processing. As noted in the Experimental Details, the 973~K anneal is included as a deliberately severe reference condition to probe this upper limit of thermal degradation, rather than as a technologically recommended processing step.

\subsection{XPS Analysis}
XPS depth profiling was employed to investigate the chemical evolution of the Ni/GeSn interface of Ni/\allowbreak Ge$_{1-x}$Sn$_x$/\allowbreak Ge/\allowbreak Gd$_2$O$_3$/\allowbreak Si(111) layered structure for both Sn concentrations (8\% and 10\%). Sequential sputtering was carried out up to 11 etching cycles, and the chemical states of Ni, Ge, and Sn were monitored across the Ni/GeSn interface and into the GeSn layer for the as-deposited and RTA-treated samples at 573 K and 973 K. The acquired core-level spectra were plotted as a function of the sputtering level (0-11), where sputtering level ~0 corresponds to the sample surface (top of the Ni layer), while increasing sputtering levels correspond to progressively deeper regions of the heterostructure, extending from the Ni layer through the Ni/GeSn interface and into the underlying GeSn layer.

\begin{figure}[t!]
	\centering
	\includegraphics[width=0.9\textwidth]{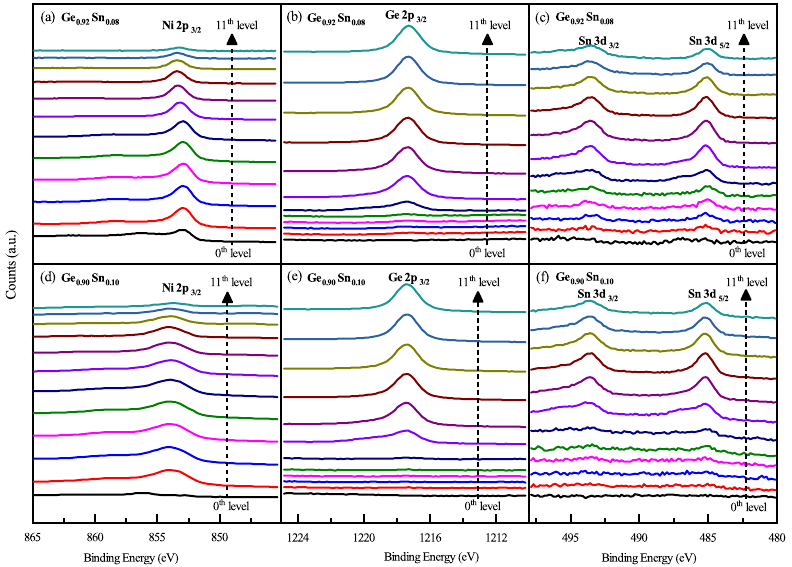}
	\caption{XPS depth profile spectra of the as-deposited Ni/\allowbreak Ge$_{1-x}$Sn$_x$/\allowbreak Ge/\allowbreak Gd$_2$O$_3$/\allowbreak Si(111) layered structure for two Sn concentrations. Panels~(a), (b), and (c) show the Ni~2p$_{3/2}$, Ge~2p$_{3/2}$ and Sn~3d spectra for Ge$_{0.92}$Sn$_{0.08}$ sample respectively. Similarly, panel (d), (e) and (f) show the Ni~2p$_{3/2}$, Ge~2p$_{3/2}$ and Sn~3d spectra for Ge$_{0.90}$Sn$_{0.10}$ sample respectively. Each spectrum represents successive sputtering cycles (0 to 11 cycle) from the surface into the film, illustrating the evolution of chemical states across the Ni/GeSn interface.}
	\label{xps1}
\end{figure}

The XPS depth profile spectra for the as-deposited Ni/Ge$_{1-x}$Sn$_{x}$ layers are presented in Fig.~\ref{xps1}. The data indicate that within the GeSn layer, the Ge 2p$_{3/2}$ spectra exhibit a dominant peak centered at 1217-1218~eV, which can be attributed to Ge atoms bonded within the GeSn alloy lattice. Similarly, the Sn 3d spectra show the characteristic Sn $3d_{5/2}$ (~484-485~eV) and Sn $3d_{3/2}$ (493-494~eV) components, confirming the incorporation of Sn into the alloy matrix \cite{bouschet2022photoelectron,braun2022oxide}. With increasing sputtering depth, the Sn intensity gradually increased beyond approximately etch cycles 6-7 and eventually approached a nearly constant value corresponding to the bulk GeSn film. Although the Ge and Sn signals began to emerge at earlier sputtering stages (around cycles 4-5), the Ni signal remained detectable until approximately cycles 9-10. This extended presence of the Ni signal does not imply a deeper physical penetration of the Ni layer, but is instead attributed to sputter-induced atomic mixing and the difference in sputtering yields between Ni and GeSn, which results in a broadening of the interface region. In addition, a slight shift of the Ge $2p_{3/2}$ peak towards a lower binding energy was observed in the vicinity of the interface for both compositions. The magnitude of this shift remains comparable for the Ge$_{0.92}$Sn$_{0.08}$ and Ge$_{0.90}$Sn$_{0.10}$ samples, suggesting that the variation in Sn concentration has a negligible effect on the Ge core-level binding energy.

\begin{figure}[t!]
	\centering
	\includegraphics[width=0.9\textwidth]{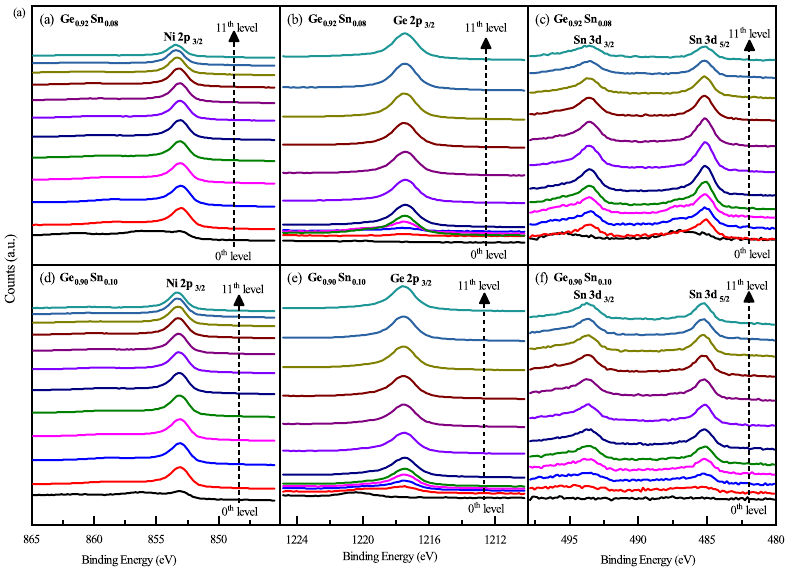}
	\caption{XPS depth profile spectra of the RTA-treated (573~K) Ni/\allowbreak Ge$_{1-x}$Sn$_x$/\allowbreak Ge/\allowbreak Gd$_2$O$_3$/\allowbreak Si(111) layered structure for two Sn concentrations. Panels~(a), (b), and (c) show the Ni~2p$_{3/2}$, Ge~2p$_{3/2}$ and Sn~3d spectra for Ge$_{0.92}$Sn$_{0.08}$ sample respectively. Similarly, panel (d), (e) and (f) show the Ni~2p$_{3/2}$, Ge~2p$_{3/2}$ and Sn~3d spectra for Ge$_{0.90}$Sn$_{0.10}$ sample respectively. Each spectrum represents successive sputtering cycles (0 to 11 cycle) from the surface into the film, illustrating the evolution of chemical states across the Ni/GeSn interface.}
	\label{xps2}
\end{figure}

The XPS depth profiles of the samples annealed at 573 K displayed in Fig.~\ref{xps2} revealed noticeable changes in all three elements, indicating the onset of the interfacial reactions. The Ni $2p_{3/2}$ intensity begins to decrease after the fourth sputtering cycle, suggesting that the interfacial boundary remains close to its initial depth. However, the reduction in Ni intensity was more gradual than that observed in the as-deposited samples, indicating that the Ni layer was no longer spatially confined to the surface region. For the Ge$_{0.92}$Sn$_{0.08}$ sample, the Ni signal decreases rapidly between sputtering cycles 4 and 6, whereas the Ge$_{0.90}$Sn$_{0.10}$ sample exhibits a more gradual reduction with detectable Ni intensity extending up to cycles 6-7. This behavior suggests broadening of the Ni/GeSn interface due to thermally induced interdiffusion and atomic mixing.

At this annealing temperature, a distinct difference in the near-surface elemental distribution emerged between the two compositions. In the Ge$_{0.92}$Sn$_{0.08}$ sample, Sn was detectable at the surface, while no Ge signal was observed. In contrast, the Ge$_{0.90}$Sn$_{0.10}$ sample exhibited the opposite behavior, where Ge was detected near the surface, while Sn was absent at the initial sputtering cycle. The distinct compositional behavior suggests different elemental redistribution mechanisms during the thermal treatment and can be correlated with the microstructural evolution observed from the pole figure measurements.

The pole figure analysis indicates that the Ge$_{0.90}$Sn$_{0.10}$ (see Fig.~\ref{PF}(e)) sample developed microtwin defects and associated stacking faults after annealing at 573 K, whereas the Ge$_{0.92}$Sn$_{0.08}$ (see Fig.~\ref{PF}(b)) sample maintained its crystallographic integrity without noticeable defect formation. Based on these observations, we propose the following mechanism to interpret the composition-dependent near-surface elemental distribution. The additional defect sites in the 10\% Sn sample may act as trapping centers for migrating Sn atoms and consequently suppress Sn diffusion toward the Ni contact during the short 10 s RTA process. As a result, Sn redistribution toward the surface region is reduced, and the interfacial reaction pathway can be modified, promoting Ni-Ge interactions and consequently enhancing Ge migration toward the surface region. This proposed mechanism is consistent with the detectable Ge signal and the absence of Sn during the initial sputtering cycle in the Ge$_{0.90}$Sn$_{0.10}$ sample. We emphasize that pole-figure measurements alone do not provide direct microscopic evidence of Sn trapping; confirmation would require direct microscopy and spatially resolved compositional mapping.

In contrast, the relatively defect-free Ge$_{0.92}$Sn$_{0.08}$ lattice provides fewer internal trapping sites for the Sn atoms. Consequently, thermally activated Sn atoms can diffuse more readily through the GeSn layer and along the grain boundaries within the overlying Ni layer, resulting in Sn segregation toward the surface region, where it became detectable at the initial sputtering cycle. The presence of segregated Sn can simultaneously suppress the upward migration of Ge, which is consistent with the absence of a detectable Ge signal at the surface~\cite{khelidj_nickel_2024}.

\begin{figure}[t!]
	\centering
	\includegraphics[width=0.9\textwidth]{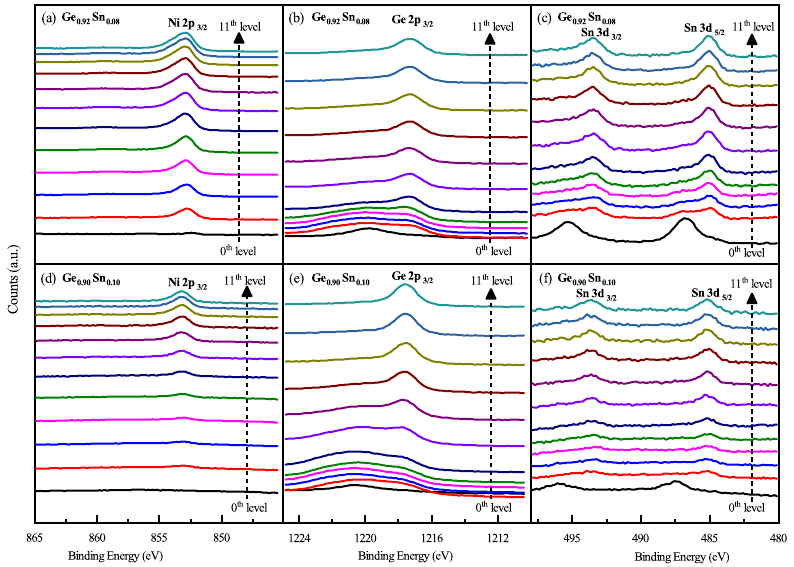}
	\caption{XPS depth profile spectra of the RTA-treated (973~K) Ni/\allowbreak Ge$_{1-x}$Sn$_x$/\allowbreak Ge/\allowbreak Gd$_2$O$_3$/\allowbreak Si(111) layered structure for two Sn concentrations. Panels~(a), (b), and (c) show the Ni~2p$_{3/2}$, Ge~2p$_{3/2}$ and Sn~3d spectra for Ge$_{0.92}$Sn$_{0.08}$ sample respectively. Similarly, panel (d), (e) and (f) show the Ni~2p$_{3/2}$, Ge~2p$_{3/2}$ and Sn~3d spectra for Ge$_{0.90}$Sn$_{0.10}$ sample respectively. Each spectrum represents successive sputtering cycles (0 to 11 cycle) from the surface into the film, illustrating the evolution of chemical states across the Ni/GeSn interface.}
	\label{xps3}
\end{figure}

Fig.~\ref{xps3} shows the XPS depth profile spectra of the Ni/Ge$_{1-x}$Sn$_x$ samples after annealing at 973~K, revealing extensive interfacial reactions and significant elemental redistribution in both compositions. Unlike the as-deposited and 573~K annealed samples, where Ge and Sn signals emerged only after sputtering cycles 4-5, strong Ge and Sn signals were detected from the initial sputtering cycle at 973~K. The simultaneous presence of Ni, Ge, and Sn throughout the sputtering profile indicates that the initially abrupt Ni/GeSn interface had evolved into a highly intermixed reaction region, demonstrating extensive interdiffusion and suggesting substantial consumption of the metallic Ni layer through Ni-Ge-Sn intermixing \cite{quintero2020}. Although this chemical intermixing is consistent with the possible formation of a ternary Ni-Ge-Sn intermetallic phase reported in the literature, the present XPS depth profiles alone do not uniquely identify a specific ternary intermetallic compound. Furthermore, the Ni $2p_{3/2}$ peak exhibits considerable broadening compared to lower annealing temperatures, indicating that Ni atoms occupy multiple chemical environments within the intermixed region, consistent with extensive intermixing and multiple bonding environments rather than a uniquely phase-identified ternary compound \cite{bouschet2022photoelectron}.

Similarly, the Ge $2p_{3/2}$ spectra exhibited significant broadening and asymmetry over an extended sputtering range, indicating strong chemical mixing and interdiffusion at the interface. The broadening is slightly more pronounced in the Ge$_{0.90}$Sn$_{0.10}$ sample compared with Ge$_{0.92}$Sn$_{0.08}$, suggesting more extensive Ni-Ge-Sn intermixing (and possibly alloy formation) at higher Sn concentrations. This observation is further supported by the pole figure measurements, which showed complete loss of crystalline order in the 10\% Sn sample after annealing at 973~K, indicating a highly disordered matrix that facilitates enhanced atomic intermixing.

The Sn $3d$ depth profiles further supported this interpretation. A strong Sn signal was observed at the initial sputtering cycle for both compositions at a binding energy higher than that of the bulk Sn $3d$ peak, which is characteristic of the oxidized Sn species (SnO$_x$). The SnO$_x$ signal decreased rapidly with increasing sputtering depth, confirming that the oxidized species were confined to the near-surface region. In contrast to the composition-dependent behavior observed at 573~K, the appearance of SnO$_x$ in both samples at 973~K suggests that the proposed defect-mediated Sn-trapping mechanism becomes ineffective at elevated temperatures. The increased thermal energy at 973~K promoted enhanced atomic mobility, allowing Sn segregation toward the surface, regardless of the initial microstructural state of the GeSn layer. The presence of Sn peaks at the zero etching level for the 973~K RTA samples confirms surface Sn segregation, whereas Ge is assumed to remain uniformly distributed without segregation, consistent with existing reports on Ni/GeSn contacts subjected to 823~K RTA \cite{quintero2020}.

\begin{figure}[ht!]
	\centering
	\includegraphics[width=0.9\textwidth]{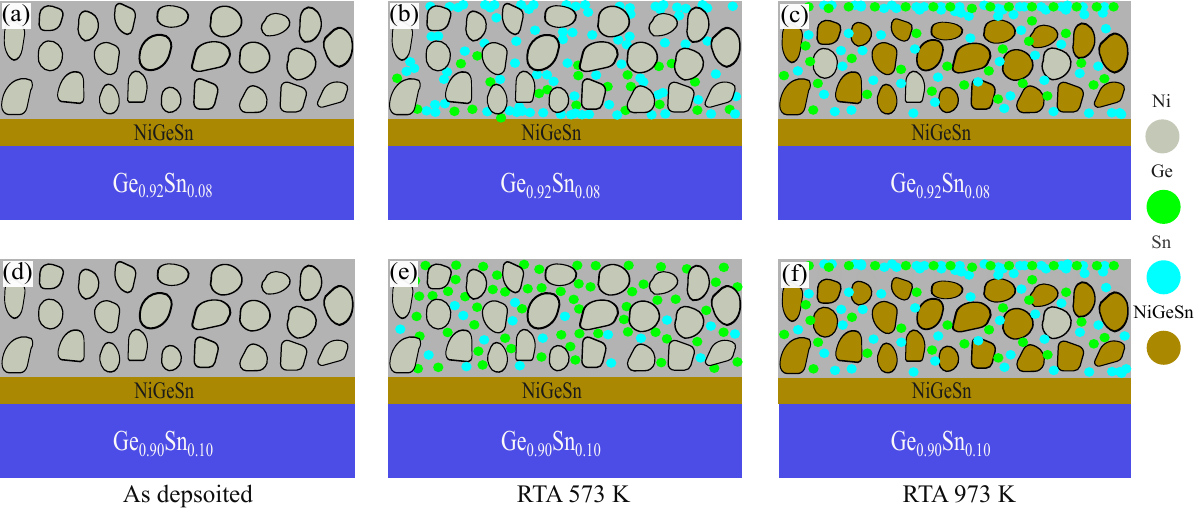}
	\caption{Schematic representation of Sn out-diffusion into the Ni layer through grain boundaries for Ge$_{0.92}$Sn$_{0.08}$ and Ge$_{0.90}$Sn$_{0.10}$ samples. Panels (a-c) correspond to the Ge$_{0.92}$Sn$_{0.08}$ sample in the as-deposited state and after RTA at 573~K and 973~K, respectively, whereas panels (d-f) correspond to the Ge$_{0.90}$Sn$_{0.10}$ sample under the same conditions.}
	\label{Sndiffschematic}
\end{figure}

Based on the XPS findings discussed above, the evolution of the Ni/Ge$_{1-x}$Sn$_x$ interface, including the interfacial diffusion and Sn segregation behavior upon RTA treatment, is schematically illustrated in Fig.~\ref{Sndiffschematic}(a-f). These schematically illustrated interfacial modifications are expected to play an important role in understanding the observed magnetization characteristics. It is reasonable to assume that the presence of Ni nanocrystallites with different size distributions within the Ni layer may contribute significantly to the observed magnetization behavior, particularly $M(T)$, and $M(H)$ data. This aspect will be further discussed in the subsequent analysis of the magnetic properties \cite{singh2025viable, Knobel2008}. Panels (a-c) correspond to the Ge$_{0.92}$Sn$_{0.08}$ sample, while panels (d-f) represent the Ge$_{0.90}$Sn$_{0.10}$ sample under as-deposited, 573~K RTA, and 973~K RTA conditions, respectively.

In the as-deposited samples presented in Fig.~\ref{Sndiffschematic}(a,d), the schematic illustrates the isolated Ni nanocrystallites distributed within the deposited Ni matrix above the GeSn layer. Consistent with the XPS results for the as-deposited samples presented in Fig.~\ref{xps1}, negligible interdiffusion occurred across the Ni/GeSn interface, with Sn remaining uniformly incorporated within the Ge host matrix at concentrations of 8\% and 10\%, thereby preserving the integrity of the GeSn layer. Consequently, the interface remained relatively sharp and chemically distinct, with no segregation effect.

The diffusion and segregation behavior of the samples upon RTA at 573~K is schematically illustrated in Fig.~\ref{Sndiffschematic}(b,e). Consistent with the XPS results presented in Fig.~\ref{xps2}, the onset of thermally activated diffusion of Sn and Ge atoms from the GeSn layer into the overlying Ni layer was observed, predominantly through the grain boundaries of the Ni nanocrystallites. Owing to the relatively low eutectic temperature of the GeSn alloy, Sn atoms exhibited enhanced mobility and diffused outward from the Ge matrix, while Ge atoms simultaneously migrated toward the Ni layer. The diffused atoms accumulated near the Ni/GeSn interface and between adjacent Ni nanocrystallites, promoting local atomic mixing and the possible initial formation of Ni-Ge-Sn intermixed regions. More importantly, at 573~K, for the 8\% Sn content sample, Sn atoms precipitated at the top of the Ni layer due to the micro-rotation-free nature of the sample, as evidenced by the pole-figure data (see Fig.~\ref{PF}(b)).

In contrast, for the 10\% Sn content sample, Sn atoms did not precipitate; instead, within the proposed mechanism, they predominantly underwent defect-mediated trapping due to the presence of micro-rotational defects (see Fig.~\ref{PF}(e)). Apart from this, due to its higher Sn concentration, the Ge$_{0.90}$Sn$_{0.10}$ sample shown in Fig.~\ref{Sndiffschematic}(e) exhibited comparatively stronger diffusion than the Ge$_{0.92}$Sn$_{0.08}$ sample shown in Fig.~\ref{Sndiffschematic}(b).

Furthermore, at a higher RTA temperature of 973~K, the XPS results presented in Fig.~\ref{xps3} indicate significantly enhanced diffusion of Sn and Ge atoms. The persistence of a relatively flat Ni $2p_{3/2}$ peak profile (see Fig.~\ref{xps3}(d)) further supports the segregation of Sn toward the topmost region of the Ni layer. Accordingly, the schematic illustrations in Fig.~\ref{Sndiffschematic}(c,f) represent the enhanced out-diffusion and segregation behavior of Sn and Ge atoms at 973~K. As schematically portrayed, a substantially larger number of Sn and Ge atoms diffused into the Ni layer and became more uniformly distributed throughout the film. The extensive intermixing among Ni, Ge, and Sn atoms is consistent with the proposed formation of Ni-Ge-Sn intermixed regions both within the Ni layer and near the interface, while Sn segregation at the surface leads to the formation of a Sn-rich top layer. Furthermore, due to its higher Sn concentration, the Ge$_{0.90}$Sn$_{0.10}$ sample shown in Fig.~\ref{Sndiffschematic}(f) exhibited comparatively stronger diffusion and segregation effects than the Ge$_{0.92}$Sn$_{0.08}$ sample shown in Fig.~\ref{Sndiffschematic}(c), resulting in a more pronounced degree of interfacial alloying and atomic redistribution.

\subsection{Magnetization studies}
Measurements of $M(T)$ and $M(H)$ are important for Ni/GeSn layered structures, because they provide insight into the effects of annealing-induced interfacial diffusion, alloying, and Sn segregation on the magnetic properties. Given the non-magnetic nature of Ge, Sn, and GeSn, the measured magnetic properties predominantly originate from the ferromagnetic Ni layer. Hence, variations in the magnetic properties can be correlated with modifications in the Ni layer induced by interfacial alloy formation near the Ni/GeSn interface, thereby providing complementary insight into interfacial evolution. As revealed by the XPS analysis, RTA processing promoted Ge and Sn out-diffusion, the formation of Ni-rich clusters and possible intermixed alloy regions, as well as Sn segregation at the top of the Ni layer. These thermally induced structural modifications make magnetic measurements valuable for correlating the interfacial evolution with the integrity of the metallic Ni layer. In particular, the analysis of $M(T)$ helps identify the magnetic state of the system, such as ferromagnetic (FM) or superparamagnetic (SPM) behavior, and determines the blocking temperature ($T_B$)~\cite{singh2025viable,roy2023effect, Mekala2016}. Therefore, any shift in $T_B$, induced by RTA, can be attributed to modifications in the local environment of the Ni nanocrystallites, including Ni/GeSn interfacial alloying, Sn segregation, and possible magnetic interactions between adjacent Ni nanocrystallites.

\subsubsection{Temperature dependent magnetization studies}
\begin{figure}[b!]
	\centering
	\includegraphics[width=0.9\textwidth]{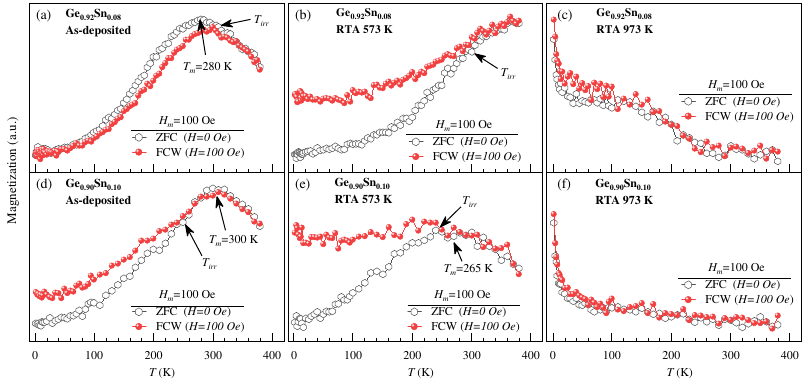}
	\caption{ZFC-FC magnetization curves measured under an in-plane field for Ni/\allowbreak Ge$_{1-x}$Sn$_x$/\allowbreak Ge/\allowbreak Gd$_2$O$_3$/\allowbreak Si(111) layered structure: (a-c) Ge$_{0.92}$Sn$_{0.08}$ in the as-deposited, RTA 573~K, and RTA 973~K conditions, respectively; (d-f) Ge$_{0.90}$Sn$_{0.10}$ under the same conditions.}
	\label{MH1}
\end{figure}

Accordingly, first magnetic behavior, $M(T)$ measurements were performed over the temperature range of 2-380~K using zero-field-cooled (ZFC) and field-cooled warming (FCW) protocols. Figures~\ref{MH1}(a-c) and \ref{MH1}(d-f) show the corresponding $M(T)$ curves for the Ge$_{0.92}$Sn$_{0.08}$ and Ge$_{0.90}$Sn$_{0.10}$ samples, respectively, under the as-deposited, RTA 573~K, and RTA 973~K conditions. The ZFC and FCW magnetization measurements were performed by cooling the samples from 380~K to 2~K in the absence of an external magnetic field and under an applied magnetic field of 100~Oe respectively. Subsequently, the ZFC and FCW magnetization data were recorded during warming from 2~K to 380~K under a small in-plane measuring magnetic field of $H_{\mathrm{m}} = 100$~Oe.

Referring to Fig.~\ref{MH1}, two characteristic temperatures, namely $T_{\mathrm{irr}}$ and $T_{\mathrm{m}}$, can be observed. Here, $T_{\mathrm{irr}}$ corresponds to the bifurcation temperature at which the ZFC and FC curves bifurcate, while $T_{\mathrm{m}}$ represents the temperature at which the ZFC magnetization reaches its maximum value~\cite{Mekala2016, Knobel2008, Vigneswari2026Fe}, commonly referred to as the blocking temperature ($T_{B}$). Apart from this, the samples annealed at an RTA of 573~K exhibited a decrease in FC magnetization below $T_{\mathrm{m}}$.

\begin{figure}[t]
	\centering
	\includegraphics[width=1\textwidth]{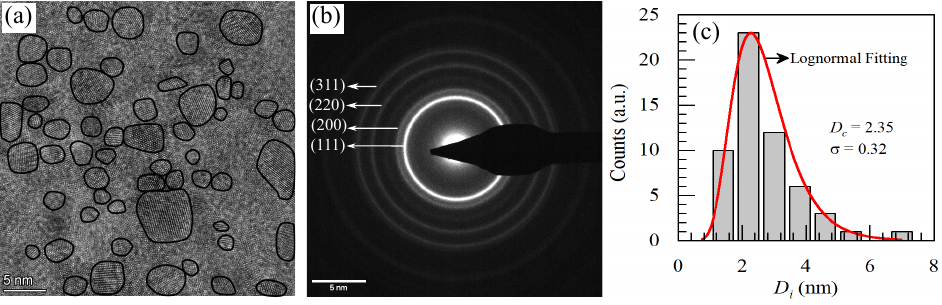}
	\caption{Direct characterization of as-deposited Ni grown with the same
		deposition parameters as in the Ni/GeSn samples.
		(a)~Plan-view TEM with nanocrystallites outlined (scale bar 5~nm).
		(b)~SAED pattern with continuous concentric rings (scale bar 5~nm).
		(c)~Crystallite diameter histogram with log-normal fit.}
	\label{Ni_cluster}
\end{figure}

The presence of the ZFC hump, $T_{\mathrm{m}}$ and $T_{\mathrm{irr}}$ in Fig.~\ref{MH1}(a,b,d,e) originates from the blocking and unblocking of magnetic moments in polycrystalline Ni nanocrystallites, similar to behavior reported for Si/Ni multilayers~\cite{singh2025viable}, Fe-doped ZnS nanoparticles~\cite{Vigneswari2026Fe}, and Dy$_x$Fe$_{100-x}$ nanogranular thin films~\cite{Mekala2016}. The ZFC hump is characteristic of the superparamagnetic behavior of the Ni nanocrystallites ~\cite{Knobel2008, singh2025viable}. To explain the origin of $T_{\mathrm{m}}$ and the broad ZFC hump, the Ni layer is taken to consist of polycrystalline Ni nanocrystallites with varying sizes, as schematically illustrated in Fig.~\ref{Sndiffschematic}. Previous studies have shown that a log-normal distribution of crystallite sizes in polycrystalline Ni films gives rise to a corresponding distribution of blocking temperatures ($T_{\mathrm{B}i}$), since each nanocrystallite possesses its own characteristic blocking temperature determined by its volume~\cite{Bedanta2009,singh2025viable,Bruvera2015,Knobel2008}. To provide direct microstructural support for this description, a Ni film was deposited onto a TEM grid under the same deposition parameters as used for the Ni/GeSn samples. The corresponding plan-view TEM, selected-area electron diffraction (SAED), and crystallite size analysis are presented in Fig.~\ref{Ni_cluster}. Panel~(a) shows discrete Ni nanocrystallites with lattice contrast and clearly resolved grain boundaries. Panel~(b) exhibits continuous concentric SAED rings, consistent with randomly oriented polycrystalline Ni nanocrystallites rather than a single-crystalline film. Panel~(c) shows the crystallite diameter histogram obtained from the TEM images, together with a log-normal fit. The crystallite size distribution was fitted using the log-normal distribution function~\cite{Knobel2008,Bedanta2009}
\begin{equation}
	\label{eq:lognormal}
	f(D)=\frac{1}{\sqrt{2\pi}\,\sigma D_i}\exp\left[-\frac{\left(\ln (D_i/ D_0)\right)^2}{2\sigma^2}\right],
\end{equation}
where $D_i$ is the crystallite diameter, $D_0$ is the median diameter, and $\sigma$ is the standard deviation of $\ln D_i$. The experimental histogram is asymmetric, with a peak near approximately 2-3~nm and a tail extending toward larger diameters (up to $\sim$6-7~nm), in good agreement with Eq.~\eqref{eq:lognormal}. Since the blocking temperature of each nanocrystallite scales with its volume ($T_B \propto KV/k_B$), the measured log-normal diameter distribution implies a corresponding distribution of blocking temperatures $T_{\mathrm{B}i}$. Consequently, the broad and asymmetric ZFC magnetization maximum observed for the as-deposited Ni/GeSn samples can be understood as a manifestation of this size distribution of Ni nanocrystallites, in line with previous reports on polycrystalline Ni films and related granular systems~\cite{Bedanta2009,singh2025viable,Knobel2008}.

During ZFC cooling from 380~K to 2~K in the absence of an external magnetic field, the magnetic moments of the Ni nanocrystallites freeze in random orientations~\cite{Bedanta2009, singh2025viable}. Consequently, at 2~K, the moments cannot align effectively with the small measuring field, $H_{\mathrm{m}}$, resulting in a low net magnetization. Upon subsequent warming, thermal energy progressively overcomes the anisotropy energy barriers and unblocks the magnetic moments according to their respective $T_{\mathrm{B}i}$. As a result, nanocrystallites with lower blocking temperatures become magnetically unblocked first and align with the applied field, while larger nanocrystallites remain blocked until higher temperatures. Consequently, the ZFC magnetization gradually increased and reached a maximum at $T_{\mathrm{m}}$ (or $T_B$), corresponding to the temperature at which the maximum number of Ni nanocrystallites became unblocked and contributed to the net magnetization~\cite{Knobel2008}. Above $T_{\mathrm{m}}$, thermal fluctuations increasingly randomize the magnetic moments, leading to a gradual decrease in the magnetization up to 380~K~\cite{Bedanta2009,Knobel2008}. Therefore, the broad ZFC hump and its transition width can be attributed to the progressive blocking and unblocking of magnetic moments in Ni nanocrystallites possessing a distribution of blocking temperatures that follows the experimentally observed log-normal crystallite size distribution in Fig.~\ref{Ni_cluster}(c).

In the FCW condition, however, the magnetic moments remain partially aligned along the cooling field, resulting in higher magnetization than the ZFC curve at low temperatures, as expected in the absence of significant magnetic interactions. This behavior gives rise to the ZFC-FCW bifurcation. Typically, in ferromagnetic systems measured under a small magnetic field $H_{\mathrm{m}} \leq 200$~Oe, the FCW magnetization continuously increases with decreasing temperature below $T_{\mathrm{m}}$ and does not exhibit a hump-like feature similar to the ZFC curve, as commonly reported for thin films, multilayers~\cite{Grundy2000,Sudhansu2014,Villacorta2011}, and nanoparticles~\cite{Vigneswari2026Co,Topkaya2012,Vigneswari2026Fe}, and/or nanostructures. However, in the present study, the decrease in FCW magnetization within the temperature range $2~\mathrm{K} \leq T \leq T_B$, resembling ZFC behavior, may originate from antiferromagnetic interactions between adjacent Ni nanocrystallites~\cite{singh2025viable} or from spin-glass-like behavior associated with frustrated magnetic interactions~\cite{Molina2012}; these scenarios are discussed as possible interpretations rather than uniquely established ground states.

For the as-deposited samples, the $T_B$ values of Ge$_{0.92}$Sn$_{0.08}$ (see Fig.~\ref{MH1}(a)) and Ge$_{0.90}$Sn$_{0.10}$ (see Fig.~\ref{MH1}(d)) are approximately 280~K and 300~K, respectively. The comparatively higher $T_{\mathrm{m}}$ of the Ge$_{0.90}$Sn$_{0.10}$ sample suggests enhanced magnetic anisotropy barriers~\cite{Bedanta2009}, stronger interparticle magnetic interactions between adjacent Ni nanocrystallites~\cite{Balaev2017}, and/or a relatively larger average nanocrystallite size, consistent with the dependence of blocking temperature on particle volume ($T_B \propto V$) \cite{singh2025viable}. However, the FCW behavior followed a similar trend for both as-deposited samples, indicating negligible interaction-driven variations in $T_B$. Apart from this, since the Ni layer thickness is nearly identical (~15 nm) in both samples, a comparable size distribution of Ni nanocrystallites is expected, which minimizes any significant shift in the ZFC hump position arising solely from nanocrystallite size effects. Therefore, the observed difference in the $T_B$ is more likely attributed to variations in magnetic anisotropy, possibly induced by differences in Sn concentration and the resulting possible Ni-Ge-Sn interfacial alloying near the Ni and GeSn interface.

Samples subjected to RTA at 573~K, in contrast, showed significant modifications in the magnetic response of the Ni layer, where $T_B$ shifted beyond 380~K for Ge$_{0.92}$Sn$_{0.08}$ (see Fig.~\ref{MH1}(b)), while it decreased to approximately 265~K for Ge$_{0.90}$Sn$_{0.10}$ (see Fig.~\ref{MH1}(e)). In addition, the FCW magnetization decreased below $T_B$ for Ge$_{0.92}$Sn$_{0.08}$, whereas it remained nearly unchanged for Ge$_{0.90}$Sn$_{0.10}$ upon warming from 2~K to $T_B$, accompanied by a clear FCW-ZFC bifurcation.

This contrasting magnetic behavior can be understood with the aid of the schematic illustrations shown in Fig.~\ref{Sndiffschematic}(b,e), which depict RTA-induced out-diffusion of Sn and Ge atoms into the Ni layer. Owing to the relatively low eutectic temperature of the GeSn alloy ($\sim$239.1$^\circ$C)~\cite{singh2025growth,gayakwad_epitaxial_2023}, both Sn and Ge atoms diffuse outward from the GeSn layer during RTA process, with Sn exhibiting comparatively higher precipitation at the surface. However, the extent of diffusion and subsequent alloying strongly depends on the crystalline quality and defect density of the GeSn layer. Within the proposed microtwin-related mechanism discussed above, for Ge$_{0.92}$Sn$_{0.08}$, the relatively superior crystalline quality and the absence of micro-rotational twinning, as evidenced by the pole figure shown in Fig.~\ref{PF}(b), would suppress defect-mediated Sn trapping. As a result, most of the Sn remained available near the top surface of the GeSn layer owing to the Sn precipitation effect (see Fig.~\ref{Sndiffschematic}(b))~\cite{singh2025growth}. Consequently, a comparatively lower amount of Ni-Ge-Sn intermixed regions is expected to form locally between adjacent Ni nanocrystallites, allowing antiferromagnetic interactions between the Ni nanocrystallites to persist compared with the as-deposited sample (Fig.~\ref{MH1}(a)). This explains the reduction in FCW magnetization below $T_B$.

In contrast, Ge$_{0.90}$Sn$_{0.10}$ exhibits a comparatively higher density of micro-rotational defects, as evidenced by the pole figure shown in Fig.~\ref{PF}(e), which is proposed to promote defect-assisted Sn trapping through segregation effect and reduces the effective Sn concentration diffusing into the Ni layer, as schematically illustrated in Fig.~\ref{Sndiffschematic}(e). The occurrence of Sn segregation in GeSn samples containing 10\% Sn after RTA at 623~K has been reported previously, providing further support for the segregation behavior observed in the present study at temperatures below 973~K \cite{Wang2020}. Owing to the overall higher Sn concentration in this sample, sufficient GeSn remained available to locally form Ni-Ge-Sn intermixed regions between neighboring Ni nanocrystallites. Such local intermixing is proposed to weaken the antiferromagnetic interaction between the Ni nanocrystallites, resulting in an enhanced FCW magnetization below $T_B$ compared to the as-deposited sample (Fig.~\ref{MH1}(d)). Furthermore, the increased presence of intermixed Ni-Ge-Sn and semiconducting content within the Ni layer weakens the interparticle magnetic coupling, thereby lowering $T_B$.

A completely different magnetic response emerged after RTA at 973~K (Fig.~\ref{MH1}(c,f)) for both 8\% and 10\% Sn content samples. As schematically illustrated in Fig.~\ref{Sndiffschematic}(c,f), the diffusion of Ge and Sn becomes significantly enhanced at this temperature, making surface segregation highly probable \cite{singh2025growth, gayakwad_epitaxial_2023, Singh2025Raman}. Extensive interdiffusion within the Ni layer is consistent with widespread Ni-Ge-Sn intermixing and possible alloy formation, accompanied by elemental precipitation at the top surface. This process substantially diminishes the schematic theme of isolated pure Ni nanocrystallites as the grains become compositionally diluted by intermixing. Consequently, antiferromagnetic coupling among neighboring Ni grains is unlikely to persist, as the extensive interdiffusion transforms the Ni layer into a compositionally diluted magnetic system. As a result, the $M(T)$ data for both samples presented in Fig.~\ref{MH1}(c,f) exhibit weak ferromagnetic behavior at low temperatures, followed by a nearly paramagnetic-like response above $280$ K, similar to that reported for Co-doped ZnS diluted magnetic semiconductor nanoparticles \cite{Vigneswari2026Co}. This suggests a reduction in the effective Curie temperature relative to Ni thin films due to alloying of the Ni layer with semiconducting Ge and Sn contents \cite{singh2025viable}.

In summary, the $M(T)$ measurements reveal that the magnetic response is primarily governed by the Ni layer but is strongly influenced by interfacial diffusion, thermal processing, and composition of the GeSn layer. While the Ni layer remains of good quality and the Ni/GeSn interface is largely preserved in the as-deposited and 573~K annealed samples, annealing at 973~K leads to a transition from blocked superparamagnetic-like behavior to a predominantly paramagnetic-like response [Fig.~\ref{MH1}(c,f)]. This evolution is consistent with substantial interdiffusion, Sn segregation, and Ni-Ge-Sn intermixing (with possible alloy formation), which dilute the magnetic Ni layer and limit its effectiveness as a thermally stable contact at elevated temperatures.

\subsubsection{Magnetic field-dependent magnetization studies}
\begin{figure}[t]
	\centering
	\includegraphics[width=1\linewidth]{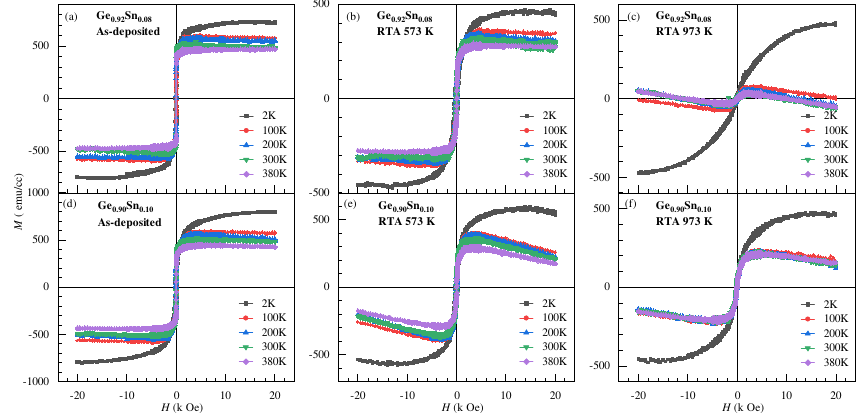}
	\caption{Full-range $M$-$H$ curves for Ni/\allowbreak Ge$_{1-x}$Sn$_x$/\allowbreak Ge/\allowbreak Gd$_2$O$_3$/\allowbreak Si(111) layered structure: (a-c) Ge$_{0.92}$Sn$_{0.08}$ in the as-deposited, RTA 573~K, and RTA 973~K conditions, respectively; (d-f) Ge$_{0.90}$Sn$_{0.10}$ under the same conditions.}
	\label{MH2}
\end{figure}

To further elucidate the origin of the observed magnetic behavior and quantify the evolution of ferromagnetism with Sn concentration and RTA effects, magnetic-field-dependent magnetization $M(H)$ measurements were performed at different temperatures (2, 100, 200, 300, and 380~K). While the ZFC-FC analysis provides insight into the temperature-dependent magnetic phases, the $M(H)$ characteristics enable direct evaluation of key magnetic parameters, including the field-dependent saturation behavior of the hysteresis loops, saturation magnetization, and coercivity. These measurements provide additional insights into the modifications of the Ni layer induced by interfacial effects and thermal processing. Fig.~\ref{MH2} shows the in-plane magnetic $M(H)$ hysteresis loops of the Ni (15~nm)/Ge$_{1-x}$Sn$_x$/Ge/Gd$_2$O$_3$/Si(111) layered structures with Sn concentrations of 8\% and 10\% under the as-deposited, RTA 573~K, and RTA 973~K conditions.

Because the sample consists of a Ni layer deposited on the substrate template, the measured magnetic moment includes contributions from both the Ni film and the substrate template. Therefore, to obtain the intrinsic magnetic response of the Ni film, the diamagnetic contribution of the substrate template (i.e. $m(T)_{\mathrm{substrate}}$) was subtracted from the total magnetic moment of the sample (i.e. $m(T)_{\mathrm{sample}}$). The substrate susceptibility $\chi_{\mathrm{substrate}} = (dm(T)/dH)_{\mathrm{substrate}}$, corresponding to the slope of the substrate $M(H)$ loop, was determined from the $M(H)$ data of the bare substrate template at a given temperature T. The corrected film magnetization at temperature $T$ was then calculated following a procedure described elsewhere~\cite{AKKERA2015}:

\begin{equation}
\label{eq:1}
M_{\mathrm{sample}}(T) = \Big(m_{\mathrm{total}}(T) - \chi_{\mathrm{substrate}} \times \left(m_{\mathrm{sample}}/m_{\mathrm{substrate}}\right) \times H\Big)/V
\end{equation}
where $m_{\mathrm{sample}}$, $m_{\mathrm{substrate}}$, and $V$ are the mass of the sample, mass of the substrate template, and volume of the 15~nm Ni layer, respectively. Here, $M_{\mathrm{sample}}(T)$ represents the sample magnetization as a function of temperature, expressed in the standard volume magnetization units of $\mathrm{emu\,cm^{-3}}$~\cite{cullity2011}.

In the as-deposited state, both compositions exhibit clear FM behavior with well-defined hysteresis and gradual low-field saturation, indicating that the magnetic response is dominated by the continuous Ni layer with only limited interfacial Ni-Ge-Sn intermixing (Fig.~\ref{Sndiffschematic}(a,d)). The gradual approach to saturation further suggested the presence of SPM contributions, in addition to FM. The saturation magnetization ($M_s$) at 2K and 300~K for the as-deposited samples, estimated from high-field extrapolation to zero field~\cite{DSINGH2024171975}, was approximately 514~emu/cc and 486~emu/cc, respectively, for both samples. These $M_s$ values at 2~K and 300~K are comparable to the reported bulk saturation magnetization values of Ni, approximately 512~emu/cc and 484~emu/cc, respectively~\cite{cullity2011,Garcia2015}.

The nearly identical $M_s$ values observed for both the as-deposited samples suggest that the extent of interfacial Ni-Ge-Sn intermixing and the remaining pure Ni layer thickness was comparable between the two samples. Furthermore, the nearly identical $M_s$ values suggest similar volume fractions, sizes, and distributions of Ni nanocrystallites in both samples. They also implied that the remaining pure Ni layer thickness exceeded 2~nm, which is consistent with the findings of Neugebauer \textit{et al.}, who reported that Ni films thicker than 2~nm exhibited $M_s$ values comparable to those of bulk Ni~\cite{Neugebauer1959}. Any intermixed interfacial region may be regarded as a magnetic dead layer, where intermixing renders a portion of the Ni layer magnetically inactive, thereby reducing the ferromagnetically active Ni thickness. However, quantifying this dead-layer thickness would require systematic thickness-dependent magnetization measurements and dedicated dead-layer modeling, which are beyond the scope of the present work~\cite{Sbiaa2026,Lau2019,Singh2022deadlayerstatic}.

After annealing at 573~K, the hysteresis loops retained their ferromagnetic nature but exhibited a noticeable reduction in saturation magnetization, accompanied by a deviation from ideal saturation, as evidenced by the gradual bending in the high-field region, particularly for the 10\% Sn-doped samples compared with the 8\% Sn-doped samples. This behavior can be understood from the schematics shown in Fig.~\ref{Sndiffschematic}(b) and (e). At 573~K, distinct diffusion behavior of Sn atoms was observed depending on the Sn concentration. In the lower Sn content sample, Sn atoms tended to precipitate near the top surface of the Ni layer. In contrast, for the sample with a higher Sn content, within the proposed mechanism, Sn atoms were predominantly trapped at micro-rotational defect sites as a result of segregation. This segregation process increased the semiconductor content distributed between Ni nanocrystallites. This effect introduced a stronger diamagnetic background contribution to the overall magnetization, represented by the modified susceptibility term, $\chi$, in Eq.~\ref{eq:2}. Subsequently, this effect led to the pronounced bending observed in the high-field $M(H)$ curves, particularly in Fig.~\ref{MH2}(e).

In contrast, RTA at 973~K for both samples resulted in significantly stronger high-field bending in the $M(H)$ loops (Fig.~\ref{MH2}(c) and (f)), indicating pronounced suppression of ferromagnetic ordering owing to extensive interdiffusion and Ni-Ge-Sn intermixing throughout the Ni layer (Fig.~\ref{Sndiffschematic}(c,f)). The dilution of ferromagnetic Ni nanocrystallites, which modifies $\chi$ in Eq.~\ref{eq:2}, bends the $M(H)$ curves in the high-field regime. The latter is supported by the high-temperature $M(T)$ data (Fig.~\ref{MH1}(c,f)), which exhibited paramagnetic behavior. At 2~K, the $M(H)$ loops remained highly unsaturated over the entire applied field range, indicating a significantly larger SPM contribution compared to the other samples. Overall, the progressive degradation of ferromagnetic hysteresis with annealing highlights the crucial role of interfacial reactions and chemical intermixing in determining the magnetic properties of Ni/GeSn system.

\subsubsection{Parsing $M(H)$ loops: FM and SPM quantification}
Furthermore, quantifying the FM and SPM contributions is essential for evaluating the evolution of the Ni layer quality during RTA, as variations in these magnetic components may sensitively reflect extensive interdiffusion and precipitation, which can lead to the formation of a paramagnetic phase in the Ni layer. Also, as shown in Fig.~\ref{MH2}, several $M(H)$ loops exhibit significant high-field bending, preventing reliable determination of the saturation magnetization ($M_s$) through extrapolation to the $y$-axis. Therefore, a quantitative separation of the FM and SPM contributions is required.

Such an analysis provides indirect insight into the structural and chemical modifications occurring at the Ni/GeSn interface, which are crucial for contact optimization after RTA exposure. Thus, to quantitatively distinguish between the FM and SPM contributions in the samples, the experimental $M(H)$ data were fitted using an appropriate theoretical model within the $\pm$4~kOe field range. Fitting was performed to examine the temperature- and annealing-dependent evolution of the magnetic components, as the measured $M(H)$ curves require a composite model comprising FM, SPM, and a linear paramagnetic (PM) background contribution. Accordingly, the total magnetization is expressed as \cite{Wan2026, Alzoubi2025, JOHN2019}:

\begin{equation}
\label{eq:2}
M(H,T) = M_{\mathrm{FM}}(H) + M_{\mathrm{SPM}}(H,T) + \chi H + C
\end{equation}
where $M_{\mathrm{FM}}$ represents the FM contribution, $M_{\mathrm{SPM}}$ is the SPM component, and where $\chi$ is the effective susceptibility, and $C$ is a constant offset. This contribution may arise from the paramagnetic phases, substrate effects, or diamagnetic background signals.

The FM component arises from magnetically ordered regions and is modeled using Preisach-type formalism \cite{Shangke2026, Moree2023}, where the total magnetization is represented as a superposition of elementary hysteretic units \cite{Mayergoyz1986}:

\begin{equation}
M_{\mathrm{FM}}(H) = \iint \mu(\alpha,\beta)\,\phi(\alpha,\beta,H)\, d\alpha\, d\beta
\label{eq:3}
\end{equation}
where $\mu(\alpha,\beta)$ is the Preisach distribution function, and $\phi(\alpha,\beta,H)$ represents bistable hysteresis operators defined by switching fields $\alpha$ and $\beta$. Such a description effectively captures the irreversible magnetization processes associated with domain-wall motion and pinning.

The SPM component arises from nanoscale magnetic clusters that do not exhibit hysteresis but instead follow a thermally activated alignment with the applied field. This contribution is described using a Langevin function, $\mathcal{L}(x)=\coth(x)-(1/x)$, given by \cite{Vidyadhar2010, JOHN2019, Alzoubi2025}:

\begin{equation}
\label{eq:4}
\begin{split}
M_{\mathrm{SPM}}(H,T) & = M_{\mathrm{SP}} \, \mathcal{L}\left(\frac{\mu H}{k_B T}\right) \\
& = M_{\mathrm{SP}}\,\left(\coth\left(\frac{\mu H}{k_B T}\right) - \left(\frac{\mu H}{k_B T}\right)^{-1}\right)
\end{split}
\end{equation}
where $M_{\mathrm{SP}}$ is the saturation magnetization of the superparamagnetic phase, $\mu$ is the average magnetic moment in terms of $\mu_B$ (Bohr magneton), and $k_B$ is the Boltzmann constant \cite{cullity2011, Vidyadhar2010}. This form is widely used to describe SPM behavior in nanoparticle systems and captures the nonlinear, reversible field dependence of magnetization. A similar modified Langevin approach has been successfully applied to Ni nanoparticle/nanograin systems exhibiting mixed magnetic phases \cite{Vidyadhar2010}.

\begin{figure}[!ht]
    \centering
    \includegraphics[width=1.0\linewidth]{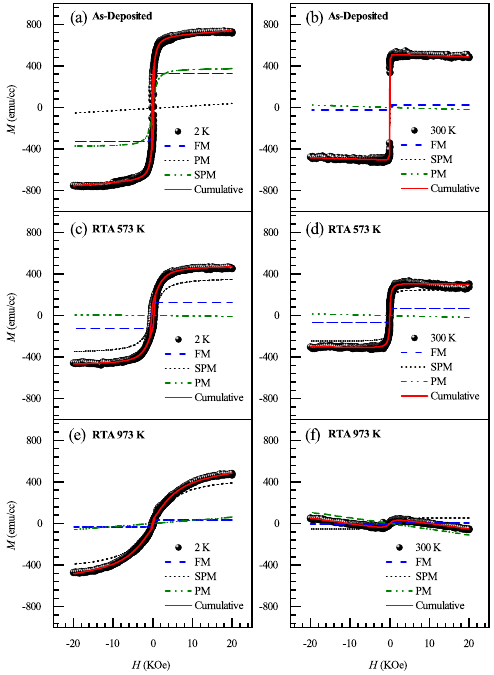}
    \caption{Experimental $M(H)$ loops of 8\% Sn content samples at 2~K, and 300~K (solid black circles) and their fitted curves using Eq.~\ref{eq:2} for Ni/\allowbreak Ge$_{0.92}$Sn$_{0.08}$/\allowbreak Ge/\allowbreak Gd$_2$O$_3$/\allowbreak Si(111) samples in the as-deposited state (a,b) and after RTA at 573 K (c,d) and 973 K (e,f).}
    \label{MH_fit_1}
\end{figure}

\begin{figure}[!ht]
    \centering
    \includegraphics[width=1.0\linewidth]{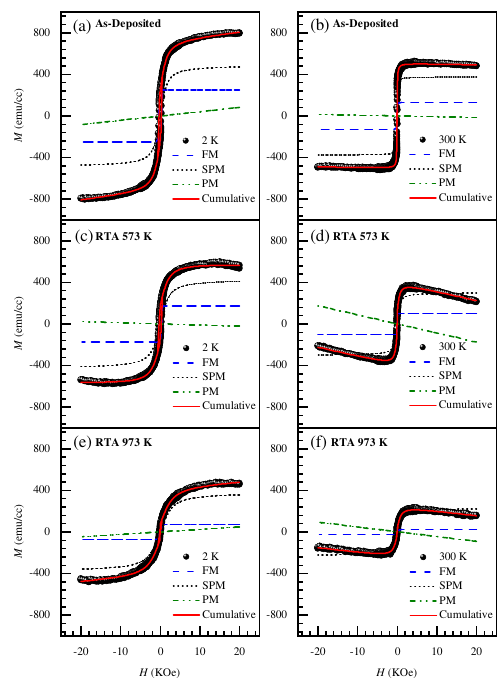}
    \caption{Experimental $M(H)$ loops of 10\% Sn content samples at 2~K, and 300~K (solid black circles) and their fitted curves using Eq.~\ref{eq:2} for Ni/\allowbreak Ge$_{0.90}$Sn$_{0.10}$/\allowbreak Ge/\allowbreak Gd$_2$O$_3$/\allowbreak Si(111) samples in the as-deposited state (a,b) and after RTA at 573 K (c,d) and 973 K (e,f).}
    \label{MH_fit_2}
\end{figure}

The experimental $M(H)$ data presented at all the measured temperatures in Fig.~\ref{MH2} were further analyzed by fitting them using Eq.~\ref{eq:2}. Using the composite model defined in Eq.~\ref{eq:2}, model parameters—including the coercive field ($H_c$), the ferromagnetic saturation magnetization ($M_{\rm S}^{\rm FM}$), and the superparamagnetic saturation magnetization ($M_{\rm SP}$)—were obtained through coordinate descent optimization. The fitted curves obtained at 2~K and 300~K for the samples with 8\% and 10\% Sn contents under the as-deposited, RTA 573~K, and RTA 973~K conditions are shown in Figs.~\ref{MH_fit_1} and \ref{MH_fit_2}, respectively. Excellent agreement between the experimental data and the fitted curves was observed for all the samples. The extracted parameters, including the coercive field ($H_c$) and the magnetization contributions ($M_{\rm S}^{\rm FM}$ and $M_{\rm SP}$), are summarized in Table~\ref{tab:FM_SPM_Hc} for each sample under different annealing conditions.
\begin{table}[t]
	\centering
	\caption{Parameters extracted from fitting the $M(H)$ data using Eq.~\ref{eq:2} for samples with 8\% and 10\% Sn contents under different annealing conditions. The table summarizes the coercive field ($H_c$), ferromagnetic saturation magnetization ($M_{\rm S}^{\rm FM}$), and superparamagnetic saturation magnetization ($M_{\rm SP}$).}
	\label{tab:FM_SPM_Hc}

	\footnotesize
	\setlength{\tabcolsep}{3.5pt}
	\renewcommand{\arraystretch}{1.05}

	\resizebox{\textwidth}{!}{%
	\begin{tabular}{llcccccc}
		\toprule
		\multirow{2}{*}{\textbf{Sample}} &
		\multirow{2}{*}{\textbf{T (K)}} &
		\multicolumn{3}{c}{\textbf{8\% Sn}} &
		\multicolumn{3}{c}{\textbf{10\% Sn}} \\

		\cmidrule(lr){3-5}
		\cmidrule(lr){6-8}

		& &
		\shortstack{\textbf{$H_c$} \\ \textbf{(Oe)}} &
		\shortstack{\textbf{$M_{\rm S}^{\rm FM}$} \\ \textbf{(emu/cm$^3$)}} &
		\shortstack{\textbf{$M_{\rm SP}$} \\ \textbf{(emu/cm$^3$)}} &
		\shortstack{\textbf{$H_c$} \\ \textbf{(Oe)}} &
		\shortstack{\textbf{$M_{\rm S}^{\rm FM}$} \\ \textbf{(emu/cm$^3$)}} &
		\shortstack{\textbf{$M_{\rm SP}$} \\ \textbf{(emu/cm$^3$)}} \\

		\midrule

		\multirow{5}{*}{As-deposited}
		& 2   & 225 & 326.5 & 380.7 & 330 & 251.0 & 487.5 \\
		& 100 & 160 & 261.0 & 362.1 & 226 & 194.5 & 427.4 \\
		& 200 & 46  & 297.9 & 271.1 & 116 & 216.1 & 366.8 \\
		& 300 & 2   & 23.1  & 484.7 & 17  & 130.7 & 374.9 \\
		& 380 & 3   & 42.7  & 414.5 & 0   & 59.4  & 371.7 \\

		\midrule

		\multirow{5}{*}{RTA 573 K}
		& 2   & 392 & 126.5 & 365.9 & 325 & 174.7 & 429.1 \\
		& 100 & 249 & 75.0  & 325.4 & 168 & 114.4 & 345.9 \\
		& 200 & 81  & 60.9  & 299.4 & 30  & 101.4 & 338.3 \\
		& 300 & 28   & 69.6  & 249.5 & 2   & 102.5 & 300.7 \\
		& 380 & 22   & 52.7  & 228.1 & 0   & 84.9  & 247.5 \\

		\midrule

		\multirow{5}{*}{RTA 973 K}
		& 2   & 206 & 33.0  & 453.0 & 148 & 70.8  & 380.0 \\
		& 100 & 161 & 12.5  & 84.0  & 98  & 33.4  & 232.4 \\
		& 200 & 5   & 9.9   & 68.1  & 64  & 33.6  & 229.6 \\
		& 300 & 2   & 3.6   & 54.5  & 44  & 25.5  & 224.6 \\
		& 380 & 0   & 4.5   & 51.8  & 28  & 28.7  & 216.8 \\

		\bottomrule
	\end{tabular}%
	}
\end{table}
It is also evident from the table that the absolute ferromagnetic saturation magnetization ($M_{\rm S}^{\rm FM}$) generally decreases with increasing temperature, and at higher temperatures (e.g., 300~K and 380~K), the superparamagnetic saturation magnetization ($M_{\rm SP}$) represents the dominant contribution to the overall magnetization for all samples. This observation is consistent with the $M(T)$ data presented in Fig.~\ref{MH1}, which suggests that the samples exhibit a stronger FM character at low temperatures, whereas the magnetic response becomes increasingly dominated by SPM behavior at elevated temperatures. The coexistence of FM and SPM behavior has also been reported in Ni/Si multilayers, where the interdiffusion of Si into the Ni layer leads to the emergence of an SPM contribution; otherwise, a pure Ni layer is expected to exhibit predominantly FM behavior \cite{singh2025viable, Singh2022deadlayerstatic}. The nearly comparable $M_{\rm S}^{\rm FM}$ and $M_{\rm SP}$ values for both samples after RTA at 573~K suggest that the Ni nanocrystallites are similar in size, which is consistent with the $M(T)$ results obtained for the 573~K RTA samples.

Additionally, at 2~K, a comparison with the as-deposited samples reveals that the FM contribution becomes almost negligible following RTA at 973~K. This behavior suggests that although Ni is widely used in metallization schemes for semiconductor systems, the substantial reduction in the FM contribution following RTA at 973~K, relative to the as-deposited state, indicates significant degradation and dilution of the metallic Ni layer. Such degradation is likely associated with extensive interdiffusion during thermal processing, leading to a pronounced dilution of the Ni layer through the incorporation of Ge and Sn atoms and the formation of Ni-Ge-Sn intermixed regions throughout the Ni layer. The RTA-induced out-diffusion and segregation of Sn from the epitaxial GeSn layer have also been reported in previous studies conducted at 923~K \cite{singh2025growth}. In contrast, the SPM contribution did not exhibit a clear or systematic trend that could be directly correlated with the degradation process.

\begin{figure}[t]
	\centering
	\includegraphics[width=0.6\linewidth]{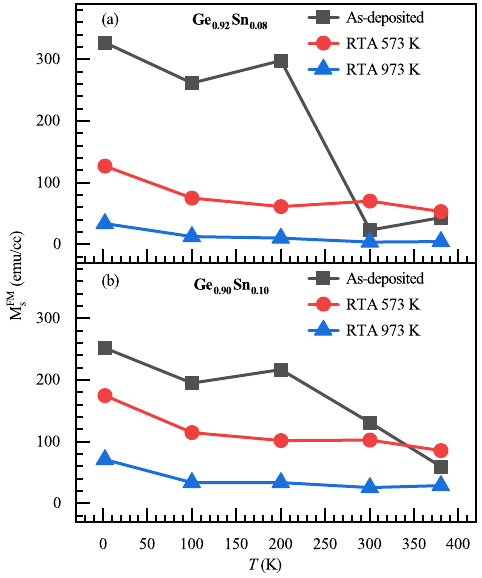}
	\caption{(a-b) Temperature dependence of the saturation magnetization ($M_{\mathrm{s}}^{\mathrm{FM}}$) for (a) Ge$_{0.92}$Sn$_{0.08}$, and (b) Ge$_{0.90}$Sn$_{0.10}$ samples in the as-deposited state (solid square symbol), and after RTA at 573~K (solid red symbol) and 973~K (solid triangle symbol).}
	\label{Ms_FM_from_fits}
\end{figure}

\begin{figure}[h]
	\centering
	\includegraphics[width=1.0\textwidth]{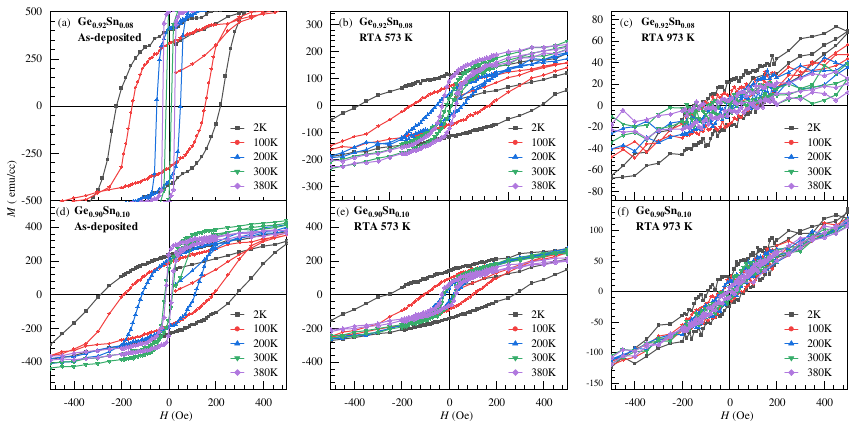}
	\caption{Low-field $M$-$H$ hysteresis loops for Ni/\allowbreak Ge$_{1-x}$Sn$_x$/\allowbreak Ge/\allowbreak Gd$_2$O$_3$/\allowbreak Si(111) layered structure: (a-c) Ge$_{0.92}$Sn$_{0.08}$ in the as-deposited, RTA 573~K, and RTA 973~K conditions, respectively; (d-f) Ge$_{0.90}$Sn$_{0.10}$ under the same conditions.}
	\label{MH4}
\end{figure}

To further elucidate the effect of RTA on the magnetic characteristics of the Ni layer, the saturation magnetization associated with the FM component ($M_{\mathrm{s}}^{\mathrm{FM}}$), extracted from the fitted FM part of the $M(H)$ data using Eq.~\ref{eq:2}, is presented in Figs.~\ref{Ms_FM_from_fits}(a) and \ref{Ms_FM_from_fits}(b) for the samples containing 8\% and 10\% Sn, respectively. For the as-deposited samples, $M_{\mathrm{s}}^{\mathrm{FM}}$ remained relatively high up to 200~K, followed by a significant decrease at higher temperatures. This behavior is consistent with the $M(T)$ data presented in Fig.~\ref{MH1}(a, d), which indicated that the magnetic response was predominantly ferromagnetic below the blocking temperature ($T_B$). As the temperature exceeds $T_B$, the contribution from SPM entities becomes increasingly significant, resulting in a reduction in $M_{\mathrm{s}}^{\mathrm{FM}}$. Consequently, the observed decrease in $M_{\mathrm{s}}^{\mathrm{FM}}$ at elevated temperatures reflects the gradual transition from predominantly ferromagnetic to superparamagnetic-dominated magnetic behavior.

Following RTA at 573~K, $M_{\mathrm{s}}^{\mathrm{FM}}$ was substantially reduced for both Sn concentrations when compared with that of the as-deposited samples. This was caused by the interdiffusion of the semiconductor material from the epitaxial GeSn layer into the Ni layer. In contrast, the samples annealed at 973~K exhibited the lowest $M_s$ values and a weak temperature dependence, indicative of a pronounced suppression of the FM contribution. This behavior is consistent with extensive interdiffusion and phase transformation at high annealing temperatures, leading to the formation of magnetically dilute or weakly magnetic phases in line with the schematic in Fig.~\ref{Sndiffschematic}.

Turning to the $H_c$ values listed in Table~\ref{tab:FM_SPM_Hc}, a clear decrease in $H_c$ was observed with increasing temperature for both the 8\% and 10\% Sn samples in the as-deposited, RTA 573~K, and RTA 973~K conditions, at which $M(H)$ loops were measured. This trend indicates a progressive weakening of the FM ordering due to thermal activation and an increasing contribution from SPM entities. Apart from this, it can be seen from Table~\ref{tab:FM_SPM_Hc} that $H_c$ increases after annealing at 573~K for the 8\% Sn sample. This behavior is consistent with reports on single-layer Co and Ni films, where an increase in $H_c$ upon annealing has been attributed to improved crystallinity and grain growth~\cite{Wang2015,Shalygina2005}. However, for the 10\% Sn sample subjected to RTA at 573~K, $H_c$ decreases rather than increases. This behavior is presumed to be associated with the greater extent of Ni-Ge-Sn alloy formation between adjacent Ni nanocrystallites due to enhanced interdiffusion, as discussed earlier in the context of the $M(T)$ data for the 10\% Sn sample after RTA at 573~K with the use of the schematic illustrated in Fig.~\ref{Sndiffschematic}(e). For RTA at 973~K, the magnetic response is qualitatively too different from that of the RTA at 573~K samples, as discussed earlier based on the data presented in Fig.~\ref{MH1}(c,f) and schematically illustrated in Fig.~\ref{Sndiffschematic}(c,f). The enhanced Ge and Sn diffusion at 973~K promotes extensive Ni-Ge-Sn intermixing, possible alloy formation, and surface segregation, leading to compositional dilution of the Ni-rich regions and reduced magnetic stability. Consequently, both the 8\% and 10\% Sn samples exhibit substantially different $H_c$ values than the corresponding RTA 573~K samples, as listed in Table~\ref{tab:FM_SPM_Hc}.

To further validate the $H_c$ values extracted from the fitting analysis, the corresponding low-field $M(H)$ hysteresis loops measured using SQUID-VSM are presented in Fig.~\ref{MH4} for samples with 8\% and 10\% Sn concentrations under the as-deposited, RTA 573~K, and RTA 973~K conditions. In the as-deposited sample (Fig.~\ref{MH4}(a,d)), the samples exhibit typical soft ferromagnetic behavior characterized by finite $H_c$, higher remanent magnetization ($M_r$) compared to RTA samples, and quick magnetic saturation (Fig.~\ref{MH2}(a,d)). Following RTA at 573 K [Fig.~\ref{MH4}(b,e)], the hysteresis loops became narrower with reduced $M_r$, indicating a weakening of ferromagnetic ordering due to interfacial reactions and the diffusion of Ge and Sn atoms into the Ni layer. After RTA at 973~K (Fig.~\ref{MH4}(c,f)), the loops became nearly reversible with negligible $H_c$ and $M_r$, signifying a substantial suppression of long-range ferromagnetic order and the predominance of thermally fluctuating magnetic entities. The close agreement between the experimentally determined and fitted $H_c$ values confirmed the robustness of the fitting approach and validated the decomposition of the $M(H)$ loops into their respective FM, SPM, and background magnetic contributions ($\chi H + C$).

Furthermore, a noteworthy observation from both Table~\ref{tab:FM_SPM_Hc} and the low-field $M(H)$ hysteresis loops in Fig.~\ref{MH4} is the rapid reduction of the $H_c$ to nearly zero at temperatures close to the peak (or hump) observed in the ZFC magnetization curves, corresponding to the $T_B$, for the as-deposited samples. This behavior is a hallmark of superparamagnetic systems, wherein $T_B$ defines the crossover between a low-temperature blocked regime, characterized by finite $H_c$ and magnetic hysteresis, and a high-temperature superparamagnetic regime, where thermal fluctuations overcome the anisotropy energy barrier \cite{singh2025viable}. This leads to a vanishing coercivity and reversible magnetization response~\cite{Bedanta2009}. For these samples, the close agreement between the temperature at which $H_c$ collapses and the experimentally determined $T_B$ supports the interpretation that the observed magnetic behavior is governed by the thermal blocking and unblocking of the superparamagnetic entities. For the Ge$_{0.92}$Sn$_{0.08}$ sample annealed at 573~K, however, the ZFC hump is broad, and the corresponding $T_B$ is expected to exceed 380~K, beyond the highest temperature probed in the present $M(H)$ measurements, while $H_c$ remains correspondingly small. This broadening, together with the small $H_c$ value, can be attributed to a broader size distribution of Ni nanocrystallites \cite{singh2025viable}, where a large fraction of the Ni nanocrystallites have already begun to unblock, rather than the point at which the full size distribution, including its higher-barrier tail, becomes unblocked. This observation is consistent with the coexistence of ferromagnetic and superparamagnetic phases inferred from the $M(H)$ fitting analysis.

Overall, the magnetic studies demonstrate that the magnetic properties of Ni/GeSn heterostructures are strongly influenced by interfacial diffusion and thermal annealing. The coexistence of the FM and SPM phases in the as-deposited samples indicates a robust metallic Ni layer, whereas extensive Ni-Ge-Sn intermixing at high annealing temperatures suppressed ferromagnetism and reduced saturation magnetization. This study establishes a direct correlation between GeSn crystalline quality, interfacial diffusion, and magnetic behavior.

\section{Conclusion}
In this work, the structural, chemical, and magnetic evolution of Ni/Ge$_{1-x}$Sn$_x$/Ge/Gd$_2$O$_3$/Si(111) heterostructures containing 8 and 10 at.\% Sn was investigated following rapid thermal annealing at 573 and 973~K. Annealing at 573~K, representing a technologically relevant thermal budget for initiating Ni-GeSn interfacial reactions, largely preserves the structural integrity of the Ge$_{0.92}$Sn$_{0.08}$ heterostructure. In contrast, the Ge$_{0.90}$Sn$_{0.10}$ sample exhibits greater crystalline degradation, including microtwin formation, with both heterostructures undergoing severe structural deterioration at 973~K, used as a deliberately severe upper-bound thermal reference.

X-ray diffraction, pole-figure analysis, and XPS depth profiling collectively reveal that increasing annealing temperature promotes Ni, Ge, and Sn interdiffusion, leading to interfacial alloying at 573~K and extensive intermixing with pronounced Sn surface segregation at 973~K. While the chemical analysis clearly establishes progressive interdiffusion, the identification of specific ternary intermetallic phases remains beyond the scope of the present study.

The magnetic response closely follows this interfacial evolution. The as-deposited samples exhibit predominantly ferromagnetic behavior characteristic of a continuous Ni layer. Annealing at 573~K modifies the balance between ferromagnetic and superparamagnetic contributions through interfacial alloying while retaining the overall magnetic functionality. In contrast, annealing at 973~K results in highly unsaturated $M(H)$ loops, a markedly reduced ferromagnetic response, and enhanced paramagnetism, reflecting extensive interfacial reactions and disruption of the metallic Ni layer.

Overall, this study establishes a direct correlation between GeSn crystalline quality, interfacial diffusion, and magnetic behavior. The comparably high thermal stability of the 8 at.\% Sn heterostructure indicates that lower Sn content favors improved structural stability of the Ni/GeSn stack during thermal processing.
\ack{TVV, DS, and KRK gratefully acknowledge the Science and Engineering Research Board (SERB), Department of Science and Technology (DST), Government of India, for funding support under Grant No.~EEQ/2022/000354. The authors also thank the Central Instrumentation Facility at IIT Gandhinagar for providing access to characterization facilities. Additionally, support from the DST-FIST program (Grant No.~SR/FST/PS-I/2020/147(C)) is acknowledged for magnetization measurements carried out using a Physical Property Measurement System (PPMS) in the Cryogen-Free Measurement System at IIT Gandhinagar.}

\section*{Declaration of competing interest}
The authors declare that they have no known competing financial interests or personal relationships that could have appeared to have influenced the work reported in this study.

\data{The data supporting the findings of this study are available from the corresponding author upon reasonable request.}

\bibliographystyle{iopart-num}
\bibliography{References}

\end{document}